\documentclass[journal,comsoc]{IEEEtran}
\IEEEoverridecommandlockouts
\usepackage{cite}
\usepackage{bm}
\usepackage{amsmath,amssymb,amsfonts}
\usepackage{graphicx}
\usepackage{multirow}
\usepackage{bm}
\usepackage{float}
\usepackage{amsmath,amsfonts}
\usepackage{algorithmic}
\usepackage{algorithm}
\usepackage{array}
\usepackage[caption=false,font=normalsize,labelfont=sf,textfont=sf]{subfig}
\usepackage{textcomp}
\usepackage{stfloats}
\usepackage{url}
\usepackage{mathtools}
\DeclarePairedDelimiter\ceil{\lceil}{\rceil}

\usepackage{verbatim}
\usepackage{cite}
\usepackage{amsmath}
\usepackage{graphicx}
\usepackage{subcaption}
\usepackage{enumitem}

\usepackage{array}
\usepackage{textcomp}
\usepackage{epstopdf}
\usepackage{caption}
\usepackage{booktabs}
\usepackage{subcaption}
\newtheorem{lemma}{Lemma}[section]
\newenvironment{proof}{ \paragraph*{\hspace{-1em}Proof}}{\hfill$\square$}

\usepackage{color,soul}
\usepackage[dvipsnames]{xcolor}
\usepackage[printonlyused,withpage]{acronym}

\acrodef{BS}{base station}
\acrodef{CSI}{channel state information}
\acrodef{ZF}{zero-forcing}
\acrodef{UE}{user equipment}
\acrodef{UL}{uplink}
\acrodef{DL}{downlink}
\acrodef{TDD}{time-division duplexing}
\acrodef{FDD}{frequency-division duplexing}
\acrodef{ZF}{zero-forzing}
\acrodef{LMMSE}{linear minimum mean squared error}
\acrodef{MRC}{maximum ratio combining}
\acrodef{BBU}{baseband unit}
\acrodef{DPD}{digital pre-distortion}
\acrodef{OTA}{over-the-air}
\acrodef{SNR}{signal to noise ratio}
\acrodef{TX}{transmit}
\acrodef{RX}{receive}
\acrodef{AWGN}{additive white Gaussian noise}
\acrodef{MIMO}{multiple-input multiple-output}
\acrodef{LS}{least-squares}
\acrodef{LIS}{large intelligent surface}
\acrodef{MVU}{minimum-variance unbiased}
\acrodef{EM}{expectation-maximization}
\acrodef{MSE}{mean square error}
\acrodef{CRLB}{Cramer-Rao Lower Bound}
\acrodef{MRC}{maximum ratio combining}
\acrodef{LOS}{line-of-sight}
\acrodef{CPL}{central perpendicular line}
\acrodef{EE}{energy efficiency}
\acrodef{AGC}{automatic gain control}
\acrodef{SNDR}{signal to noise plus distortion ratio}

\begin{document}
\title{Large Intelligent Surfaces with Low-End Receivers: From Scaling to Antenna and Panel Selection
}
\author{\IEEEauthorblockN{Ashkan Sheikhi,~\IEEEmembership{Member,~IEEE,} Juan Vidal Alegría,~\IEEEmembership{Member,~IEEE,} and Ove Edfors,~\IEEEmembership{Senior Member,~IEEE}.}
\thanks{This work was supported by  Ericsson AB, Lund, Sweden, and "SSF Large Intelligent Surfaces - Architecture and Hardware" Project CHI19-0001.}
\thanks{Ashkan Sheikhi, Juan Vidal Alegr\'{i}a, and Ove Edfors are with Department of Electrical and Information Technology, Lund University, Lund, Sweden (Email: \{ashkan.sheikhi,juan.vidal\_alegria,ove.edfors\}@eit.lth.se).}
}
\maketitle

\begin{abstract}
Feasibility of the promising large intelligent surface (LIS) concept, as well as its scalability, relies on the use of low-cost hardware components, raising concerns about the effects of hardware distortion. We analyze LIS systems with receive-chain (RX-chain) hardware distortion, showing how it may limit performance gains when scaling up these systems. In particular, using the memory-less polynomial model, analytical expressions are derived for the signal to noise plus distortion ratio (SNDR) after applying maximum ratio combining (MRC). We also study the effect of back-off and automatic gain control on the RX-chains. The derived expressions enable us to evaluate the scalability of LIS when hardware impairments are present. The cost of assuming ideal hardware is further analyzed by quantifying the minimum scaling required to achieve the same performance with non-ideal hardware. The analytical expressions derived in this work are also used to propose practical antenna selection schemes for LIS, and we show that such schemes can improve the performance significantly leading to increased energy efficiency. Specifically, by turning off RX-chains with lower contribution to the post-MRC SNDR, we can reduce the energy consumption while maintaining performance. We also consider a more practical scenario where the LIS is deployed as a grid of multi-antenna panels, and we propose panel selection schemes to optimize the complexity-performance trade-offs and improve the system overall efficiency.
\end{abstract}


\section{Introduction}
\IEEEPARstart{L}{arge} Intelligent Surfaces (LISs) have emerged as one of the next major steps in the future development of \ac{MIMO} communication systems \cite{ShaHuBeyond1,HushaPos,ShaHuHolo}. While \acp{LIS} are sometimes considered to be a scaled up version of the widely known massive \ac{MIMO} systems, they have some unique properties which distinguishes them from the massive \ac{MIMO} systems. For instance, their physical size can be compared to the user-equipment (UE) distance to the \ac{LIS}, which introduces near-field effects in the wireless channel. As a result, the common assumptions about the massive \ac{MIMO} channels are no longer valid and new effects emerge. This requires different channel models, and more investigation into the already known facts about the transmit and receive schemes in massive \ac{MIMO} systems \cite{EmilNear}. The superiority of \ac{LIS} in exploiting the available spatial degrees of freedom in the environment has been established in \cite{ShaHuBeyond1}, for the \ac{LOS} scenario, and further confirmed by the results in \cite{Pizzo_LIS,LIS_sampled} for more general propagation environments.

The \ac{LIS} technology was initially introduced in \cite{ShaHuBeyond1} as a large continuous surface of electromagnetically active material in a \ac{LOS} channel model. While having \ac{LIS} as a continuous surface is useful to conceptually understand the system, it does not seem realistically implementable with current technology in the near future. Instead, a dense antenna array deployed throughout a large surface is more realistic, which can also be interpreted as a sampled version of the continuous \ac{LIS} model. It has been shown that, if sampling is dense enough, specifically half-wavelength spatial sampling, the continuous and sampled versions of LIS have the same performance in terms of available spatial degrees of freedom due to the spatially-band-limited nature of the channels \cite{ShaHuBeyond1,LIS_sampled}. More practical implementations of \ac{LIS} consider dividing the \ac{LIS} surface into dense multi-antenna panels with a lower number of antennas, which is a more flexible and scalable design with reduced system complexity \cite{Andreia}. What we refer to as \ac{LIS} in this manuscript, corresponds to a \ac{BS} technology consisting of an active sampled surface, which should not be confused with a reconfigurable intelligent surface (RIS) or an intelligent reflecting surface (IRS).

The deployment of \acp{LIS} is envisioned to be much more challenging than conventional \ac{BS} technology because of the enormous leap in the number of transceiver chains, and the associated processing complexities. While this is motivated by the compelling gains in terms of multiplexing and beamforming performance through increasing spatial resolution \cite{MassiveMag}, the cost-efficiency of the whole system can be a great challenge due to unreasonable implementation overheads associated with deploying hundreds to thousands of transceiver chains \cite{ScaleRusek}. This may force system designers to consider the use of inexpensive hardware components in each of these transceiver chains \cite{JuanHWI}, which can introduce distortion at the output of the transceivers. Therefore, it is of high importance to study and mitigate the effects of hardware distortion in transceiver chains when scaling up \ac{LIS} to allow the use of cost-efficient transceiver chains.

There has been previous works on the design and optimization of RIS/IRS-based systems with hardware distortion \cite{NewRefIRS,IRS1,IRS2,RIS1}. For example, \cite{IRS1} investigates channel estimation in IRS-assisted MISO systems under imperfect hardware, which is modeled by additive distortion. \cite{IRS2} analyzes the achievable rate of IRS-assisted wireless systems considering hardware impairments at both the transceivers and the reflecting surface, also modeled by additive distortion. It further proposes phase shift optimization strategies that mitigate these impairments. A complete survey on these approaches can be found in \cite{NewRefIRS}. However, the literature on RIS/IRS with hardware imperfections has no direct applicability to our study due to the intrinsically different system model associated to passive reflective surfaces compared to active transceiver technologies as \ac{LIS}, which is the main focus of this work.

Transceiver AFEs non-linearity is considered to be one of the main sources of hardware distortion in receivers \cite{9481945,7817856,MollenITC2018}. It has been shown that deploying AFEs with less non-linearity may lead to major leaps in power consumption \cite{Muris2021}. Therefore, to deal with the challenge of implementing \ac{LIS} in a cost- and energy-efficient manner, it is of great importance to adopt specific schemes to compensate the effect of employing highly non-linear AFEs. A potential path to follow is to optimize the signal processing schemes and system design while maintaining the hardware quality at a minimum level. In other words, we want to get the most gain from each transceiver chain while limiting the implementation costs, e.g., by using inexpensive hardware components with limited power consumption. In \cite{AshkanICC2021} and \cite{AshkanWCNC2023}, approaches were proposed to address such challenges in massive \ac{MIMO} systems, mainly by optimizing the per-antenna digital pre-distortion (DPD) resources. Another approach is to implement antenna selection schemes and only use a portion of the array for transmission and reception. By doing so, we only use antennas which have a significant contribution to the performance and thereby increase energy efficiency.

In this paper, we focus on optimizing the performance of \ac{LIS} with non-ideal RX-chains. We first analyze the \ac{SNDR} with the purpose of studying the scaling behavior and asymptotic limits of \ac{LIS} with hardware distortion. Then, we propose receive antenna selection schemes for a \ac{LIS} with non-ideal AFEs. Optimization problems are defined and solved to illustrate the importance of performing antenna selection when scaling up \ac{LIS}. We show that selecting antennas with the strongest channels is not always the optimal solution. We then focus on more practical cases where the \ac{LIS} is implemented as a grid of panels, and transform the antenna selection problem into a panel selection problem. Low-complexity closed-form sub-optimal solutions are proposed for the panel selection. We also show that, by adopting such antenna selection and panel selection schemes, we can improve the system performance significantly for a fixed receive-chain hardware quality.

\subsection{Contributions}
The contributions of this paper are listed as follows.

\begin{itemize}
    \item We provide a framework to study the hardware distortion effect on \ac{LIS} performance while considering high-complexity non-linear polynomial model at RX-chains.
    \item We derive close-form expressions for the \ac{SNDR} of the \ac{LIS} under polynomial hardware distortion model to enable scaling and asymptotic analysis.
    \item We introduce antenna selection problems in \ac{LIS} with hardware distortion and illustrate the achievable gains of adopting optimal antenna selection. We also consider the the more practical case with a panel-based \ac{LIS} and propose closed-form solutions for panel selection with hardware distortion effects.
\end{itemize}

\subsection{Paper Outline}
The rest of this paper is organized as follows. In Section \ref{SystemModel}, the system model is introduced, which includes the \ac{LIS} deployment configurations, channel model, and hardware distortion model. In Section \ref{SNDR_Characterization}, we characterize the \ac{SNDR} and propose closed-form expressions to evaluate the \ac{LIS} performance, which is exploited for asymptotic and scaling analysis. Sections \ref{Sec_Antenna_Selection} and \ref{Sec_Panel_Selection} introduce the antenna and panel selection problems, respectively, and some proposed solutions are presented for the corresponding \ac{LIS} scenarios. In Section \ref{Sec_Numerical}, numerical results are used to further illustrate the results from previous sections. Finally, Section \ref{Sec_Conclusion} concludes the paper results. 

\subsection{Notation}
Matrices, vectors, and scalars are denoted by boldface uppercase, boldface lowercase, and italic letters, respectively. For a vector $\bm{a}$, conjugate transpose, transpose, Euclidean norm, and the $i$'th element of $\bm{a}$ are represented by $\bm{a}^H$, $\bm{a}^T$, $\|\tilde{\bm{a}}\|$, and $a_{i}$, respectively. For a scalar $a$, the complex conjugate is denoted by $\bar{a}$. We indicate zero-mean circularly-symmetric complex Gaussian random vectors with covariance matrix $\bm{C}$ as $\bm{a}\sim\mathcal{CN}(\bm{0},\bm{C})$ and the expectation of a random variable by $\mathbb{E}\{\cdot\}$. We denote a diagonal matrix with elements ${a_1,...,a_N}$ on the main diagonal by $\textnormal{diag}\left({a_1,...,a_N}\right)$ and the $N \times N$ identity matrix by $\bm{I}_N$. 

\section{System Model} \label{SystemModel}
\begin{figure}[t]
	\centering
	\includegraphics[width=2in]{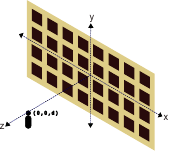}
	\caption{\ac{LIS} and UE configuration. The \ac{LIS} is centered around origin and the \ac{UE} is on the bore-sight of the \ac{LIS}.}
	\label{ArrayConfig}
\end{figure}
We consider an uplink scenario where an active \ac{LIS} consisting of $N\gg 1$ antenna elements serves a single-antenna \ac{UE}, which transmits the complex base-band symbol $s \in \mathbb{C}$ with $\mathbb{E}\{|s|^2\}=P$, over a channel $\bm{h} \in \mathbb{C}^{N\times1}$. Each receive antenna is equipped with a non-ideal RX-chain, introducing distortion into the received signal. The received signal after going through the RX-chains is expressed as
\begin{align}\label{rSystenModel}
    \bm{r}=f\left(\bm{h}s\right)+\bm{n},    
\end{align}
where $f(\cdot): \mathcal{C}^{N\times1} \rightarrow \mathcal{C}^{N\times1}$ models the overall non-ideal effects at the receiver AFEs and DFEs of the \ac{LIS}, and $\bm{n}\sim\mathcal{CN}(\bm{0},\sigma^2\bm{I}_N)$ is the additive noise.

The \ac{LIS} is assumed to span the $x$-$y$ plane with its center located at the origin, and the \ac{UE} is assumed to be located at $(0,0,d)$, where $d$ denotes the distance of the \ac{UE} to the center of the \ac{LIS}. Fig.~\ref{ArrayConfig} illustrates the configuration of the \ac{LIS} and the UE. The \ac{UE} is assumed in the far-field region of each antenna element, i.e. $d>2D_a^2/\lambda$, where $D_a$ is the maximum length of individual antenna elements. However, since the number of antenna elements can grow extremely large, the \ac{UE} may be enter the near-field region of the \ac{LIS} \cite{EmilNear,HuShaNear2024}, i.e. $d<2\bar{D}^2/\lambda$, where $\bar{D}$ is the largest distance between two antenna elements on the array. Therefore, it is important to consider both phase and amplitude variations across the array, and consider near-field effects in the channel model.

In, \cite{EmilNear}, the effect of considering near-field properties in the channel model is studied, including the variation of distance from each antenna element to the UE, different effective area, and different polarization losses due to different angle of arrivals. The results in \cite{EmilNear} imply that only considering the first two effects or considering all the three effects , even if the number of antenna elements grows extremely large, have negligible difference. Therefore, we only consider the first two effects and adopt the near-field \ac{LOS} channel model as considered in \cite{ShaHuBeyond1}. Note that in \ac{LIS} systems, the probability of having a strong \ac{LOS} component is higher than in conventional massive \ac{MIMO} due to the significantly larger aperture covered by these systems \cite{ShaHuBeyond1}. This effect is even further pronounced at higher frequencies, where reflected waves suffer greater attenuation \cite{LOS_mmWave,LOS_THz}. The amplitude of the narrow-band channel $h(x,y)=|h(x,y)|~\text{exp}\left(-j\phi(x,y)\right)$ between the UE and a point on the \ac{LIS} with coordinates $(x,y,0)$ may then be expressed as
\begin{align}
	|h(x,y)|=\frac{\sqrt{d}}{2\sqrt{\pi}(d^2+x^2+y^2)^{3/4}},
\end{align}
and the phase based on propagation delay as
\begin{align}	\phi(x,y)=2\pi\frac{\sqrt{d^2+x^2+y^2}}{\lambda},
\end{align}
where $\lambda$ is the wavelength. For the $n$'th antenna element, located at $(x_n,y_n,0)$ with effective area $A$ small enough such that $|h(x,y)|^2$ is almost constant throughout the area,\footnote{This assumption is valid in most practical \ac{MIMO} scenarios since the effective area is in the orders of $\frac{\lambda^2}{4}$ and the \ac{UE} is in the far-field of each individual antenna element, i.e. $d \gg \frac{\lambda}{2}$.} the channel gain is $|h_n|^2=A|h(x_n,y_n)|^2$.

\begin{figure}[t]
	\centering
	\includegraphics[width=3in]{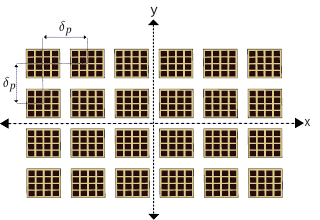}
	\caption{Panels configuration, the panel-based \ac{LIS} is centered around origin with a fixed distance $\delta_p$ between adjacent panels.}
	\label{LIS_Panels}
\end{figure}

While deploying \ac{LIS} as a co-located array of antennas is a common vision for future of wireless networks \cite{Harsh}, it may not be favorable in terms of flexibility and scalability \cite{Jesus2022}. A potentially more favorable option, without losing practicality and cost efficiency is to implement the \ac{LIS} as grid of panels, each equipped with a smaller number of equispaced antenna elements \cite{Andreia}. Fig. \ref{LIS_Panels} illustrates an example of panel configuration of a \ac{LIS} on the $x$-$y$ plane.

\textit{Remark:} 
   {This work focuses on a single-\ac{UE} scenario so as to isolate the impact of hardware distortion in LIS systems, facilitating a clearer understanding of this particular effect. However, the main results can be extended to multi-\ac{UE} scenarios, specially in hardware-distortion-limited cases where they may be directly applied. For example, in the case of multi-user scenarios with sufficiently spaced users, each of the \acp{UE} may be associated to non-overlapping active LIS areas around the respective \acp{CPL}, which is still reasonable given the large dimensionality of LIS systems.} 

\subsection{Hardware distortion model}
There are several models to opt for the distortion function $f(\cdot)$ in \eqref{rSystenModel} and the model selection is always a trade-off between accuracy and analytical tractability. One of the common models in the literature is the memory-less polynomial, which has an high accuracy in most cases, at the cost of high complexity in the analytical results. We mainly focus on this model throughout this paper to guarantee the accuracy of the analysis. In this model, the input-output relation of an RX-chain for a complex input $x_\text{in} \in \mathbb{C}$ is given by
\begin{align}\label{memoryLessModel}
	y_\text{out} = f(x_\text{in}) = \sum_{k=0}^{L-1} a_{2k+1}~x_\text{in}|x_\text{in}|^{2k},
\end{align}
where $a_{2k+1} \in \mathbb{C}$ are the model parameters, typically estimated based on input-output measurements on the RX-chains. This model only considers odd orders since they are the main source of in-band distortion \cite{DemirChannel}. For an ideal RX-chain, we have $L=1$ and $a_1=1$.

Let us first analyze the distortion behavior for a single RX-chain. To isolate the distortion from the desired signal, we can rewrite \eqref{memoryLessModel} by using the \ac{LMMSE} of $y$ given $x$ as
\begin{align}\label{BussgangEq}
	y_\text{out} = \frac{C_{y_\text{out}x_\text{in}}}{C_{x_\text{in}}} x_\text{in} + \eta,
\end{align}
where $C_{y_\text{out}x_\text{in}}$ is the covariance between $y_\text{out}$ and $x_\text{in}$, $C_{x_\text{in}}$ is the auto-correlation of $x_\text{in}$, and $\eta$ is the estimation error. Equation \eqref{BussgangEq} also corresponds to the Bussgang decomposition, which is a popular tool in hardware distortion analysis \cite{DemirBussgang}. By defining $g = {C_{y_\text{out}x_\text{in}}}/{C_{x_\text{in}}}$ as the Bussgang gain and noting that $\eta$ is uncorrelated with $x$, the auto-correlation of $\eta$ is given by 
\begin{align}\label{Cetaeta}
	C_{\eta} = C_{y_\text{out}} - |g|^2C_{x_\text{in}},
\end{align}
which can be calculated using the following lemma.

\begin{lemma}\label{BussgangLemma}
    Given the memory-less polynomial model in \eqref{memoryLessModel}, together with its Bussgang equivalent form, and assuming a Gaussian input distribution, i.e. $x_\text{in}\sim\mathcal{CN}(0,P)$, we can calculate the Bussgang gain as
   \begin{align}
        g = \frac{\mathbb{E}\left[y_\text{in}\bar{x}_\text{in}\right]}{\mathbb{E}\left[|x_\text{in}|^2\right]}=\sum_{k=0}^{L-1}a_{2k+1}(k+1)!P^k, \label{BussgangGC}
    \end{align} 
    and the distortion power as
    \begin{align}
        C_{\eta}=\mathbb{E}\left[|\eta|^2\right]=\sum_{k=1}^{2L-1}\left(k!P^k\sum_{i=1}^{k}a_{2i-1}\bar{a}_{2k-2i+1}\right)-|g|^2P, \label{BussgangCetaeta}
    \end{align} 
    where $P=\mathbb{E}\left[|x_\text{in}|^2\right]$. \\
    \begin{proof}
        See Appendix \ref{ProofsAppendix}.
    \end{proof}
\end{lemma}

In practice, the 3rd order non-linearity term, $a_{3}$, is the dominant source of RF-hardware distortions \cite{DemirChannel,schenk2008rf}. For the special case of 3rd order non-linearity, i.e. $L=2$, Lemma \eqref{BussgangLemma} gives
\begin{align}
	g=a_1+2a_3P, \label{3rdg}
\end{align}
and
\begin{align}
	C_{\eta}=2|a_3|^2P^3 \label{3rdGKappa}.
\end{align}

\subsection{\Ac{AGC} and back-off}
As mentioned earlier, the model coefficients are estimated based on input-output measurements of the RX-chain. The results are usually reported for a normalized input with unit maximum amplitude \cite{Ericsson2016}. Given the estimated parameters $\hat{a}_{2k+1}$, for the normalized input, if the actual input to the receiver has an amplitude range of $0<|{x}|<V_{max}$, we can use the following conversion to adapt the normalization
\begin{align} \label{coeffsNorm}
	a_{2k+1} = \frac{\hat{a}_{2k+1}}{(V_\text{max}^{2})^k}.
\end{align}
The value of $V_\text{max}$ can be controlled in the receiver by applying back-off to its input, either by using attenuator or by performing gain control schemes in the RX-chain amplifiers. The back-off is applied to prevent clipping, which may cause severe hardware distortion at the receivers. In our system, we define $V_\text{max}^{2}=b_\text{off}P_\text{max}$, where $b_\text{off}$ is the fixed back-off factor and $P_\text{max}$ depends on the gain control scheme at the receiver. In a typical \ac{LIS} scenario, the received power at each antenna element can have a high dynamic range, which makes the interplay between the back-off, gain control, and hardware distortion of great importance.

The back-off value, $b_\text{off}$, is a design parameter which can be selected based on the dynamic range of the received power across the \ac{LIS}. The gain control term, $P_\text{max}$, on the other hand can be set either to a fixed value for all the antenna elements, or depending on the received power on each antenna, which may be achieved by introducing per-antenna \ac{AGC} in the amplifiers. The latter case may be of less interest from a practical point of view in \ac{LIS} scenarios since there are hundreds to thousands of RX-chains across the \ac{LIS}, with a high variation of the received power. Therefore, having a perfect gain control unit on each antenna element can contribute to the already high complexity of the whole system, apart from increasing implementation costs. On the other hand, applying a fixed gain control is simpler and less expensive to implement, but has the disadvantage of excessive gain reduction on some antennas, which can reduce the energy efficiency of the amplifiers. In this paper we study both cases and compare their performance.

If all RX-chains are capable of performing individual \ac{AGC}, the per-antenna coefficients $a_{k,n}$ become
\begin{align}\label{coeffsNormAGC}
    a_{2k+1,n} = \frac{\hat{a}_{2k+1}}{(b_\text{off}P\left|{h}_n\right|^2)^k},
\end{align}
and, in this case, calculating the Bussgang parameters in Lemma \ref{BussgangLemma} results in a fixed Bussgang gain
\begin{align}
    \Tilde{g}=\sum_{k=0}^{L-1}\frac{\hat{a}_{2k+1}}{b_\text{off}^k}(k+1)! \label{BussgangAGCg0}
\end{align}
for all antenna elements. The distortion power, on the other hand, is a linearly increasing function of the input power to the corresponding antenna, i.e. $C_n=\kappa\left|{h}_n\right|^2P$, where
\begin{align}
    \kappa=\sum_{k=1}^{2L-1}\left(\frac{k!}{b_\text{off}^{k-1}}~\sum_{i=1}^{k}\hat{a}_{2i-1}\bar{\hat{a}}_{2k-2i+1}\right) - \left|\Tilde{g}\right|^2.\label{BussgangAGCDist}
\end{align}
We may thus note that by assuming per-antenna perfect \ac{AGC}, the Bussgang gain and the distortion growth rate are independent of the input power or the antenna index across the \ac{LIS}.

As we will see in the next section, the assumption of per-antenna \ac{AGC} can reduce the complexity of the analysis. Interestingly, the linear growth rate in \eqref{BussgangAGCDist} can be seen as a bridge from the memory-less polynomial model in \eqref{memoryLessModel} to the conventional additive linear distortion model, which is widely used in literature \cite{Emil2014}. One should note that the additive linear distortion model is a rough and simple approach of modeling the hardware distortion, where the single parameter $\kappa$ denotes the severeness of the hardware distortion and can be interpreted as a measure of RX-chain hardware quality.
In general, perfect per-antenna \ac{AGC} assumption and the additive linear model are not accurate. Such assumptions imply that, no matter what the input power is, the Bussgang gain, $\Tilde{g}$, and the distortion growth rate, $\kappa$, are constants. Nevertheless, we consider both the case where per-antenna \ac{AGC} is employed, as well as the case with fixed gain control, for completeness and to gain better understanding of the impact of such assumptions.

\section{SNDR Characterization} \label{SNDR_Characterization}
By using the same technique as in \eqref{BussgangEq} which is based on \ac{LMMSE} or Bussgang decomposition, the received signal in \eqref{rSystenModel} may be expressed as
\begin{align}
	&\bm{r}=\bm{G}\bm{h}s+\bm{\eta}+\bm{n},
\end{align}
where $\bm{G}=\text{diag}\left(g_1,\dots,g_N\right)$, with $g_n$ corresponding to the Busssgang gain for the $n$'th antenna as given in \eqref{BussgangGC}, and $\bm{\eta}$ is the vector containing additive distortion at the \ac{LIS}, with covariance matrix $\bm{C_{\eta\eta}}$. In general, $\bm{C_{\eta\eta}}$ is a non-diagonal matrix. However, the distortion correlation effect among antennas can be neglected in some cases, resulting in over-estimating the \ac{SNDR}, which is often considered in the literature \cite{EmilCorrNeg}. The effect of disregarding the distortion correlation is studied in Section \ref{DistortionCorrelation}.

The \ac{LIS} applies a combing vector $\bm{v}$ to the received signal $\bm{r}$ to equalize the transmitted signal. Maximum Ratio Combining (MRC) is an attractive option in \ac{LIS} scenarios because of its simplicity and reasonable performance, since it still allows us to exploit the available spatial degrees of freedom\cite{ShaHuBeyond1}. In the single-UE case without hardware distortion, MRC further provides the optimal equalization approach in terms of capacity, since it maximizes the post-processed SNR. In our scenario, the MRC vector may be expressed as $\bm{v}^T=\tilde{\bm{h}}^H/\|\tilde{\bm{h}}\|$ where $\tilde{\bm{h}}=\bm{G}\bm{h}$, can be interpreted as the effective channel, which includes the physical channel and the multiplicative hardware distortion effect. In fact, the \ac{LIS} would only be able to estimate the effective channel $\tilde{\bm{h}}$ from the uplink UE pilots since the signals received during channel estimation are also affected by the hardware distortion. We assume that the \ac{LIS} has a perfect estimate of $\tilde{\bm{h}}$ so that we can isolate the effect of hardware distortion from that of channel estimation imperfections. Note that the estimation error would only correspond to an uncorrelated additive term, which would effectively not affect our analysis on hardware distortion.

By applying the combining vector $\bm{v}^T=\tilde{\bm{h}}^H/\|\tilde{\bm{h}}\|$ to $\bm{r}$, we reach
\begin{align} \label{sndr_mrc}
\bm{v}^T\bm{r}=\frac{\tilde{\bm{h}}^H\tilde{\bm{h}}s+\tilde{\bm{h}}^H\bm{\eta}+\tilde{\bm{h}}^H\bm{n}}{\|\tilde{\bm{h}}\|}.
\end{align}
The signal, distortion, and noise power can be calculated from \eqref{sndr_mrc} considering that the distortion term is uncorrelated with the signal and has covariance $\bm{C_{\eta\eta}}$. The instantaneous \ac{SNDR} is therefore given by
\begin{align}\label{SNDRMISO_corr}
    \gamma = \frac{P\tilde{\bm{h}}^H\tilde{\bm{h}}}{\frac{\tilde{\bm{h}}^H\bm{C_{\eta\eta}}\tilde{\bm{h}}}{\tilde{\bm{h}}^H\tilde{\bm{h}}}+\sigma^2}.
\end{align}

If we assume that the distortion correlation among the antennas are negligible, i.e., $\bm{C_{\eta\eta}}$ is diagonal, the instantaneous \ac{SNDR} can be simplified, which results in more analytical tractability for studying the impact of hardware distortion on \ac{LIS} systems. We analyze the effect of this assumption in Section~\ref{DistortionCorrelation} and show that it does not invalidate the conclusions drawn regarding scaling up of \ac{LIS} systems with hardware distortion. Moreover, it leads to significant simplification of the proposed methods, at the cost of only a minor performance loss. Assuming diagonal $\bm{C_{\eta\eta}}$, i.e., $\bm{C_{\eta\eta}}=\text{diag}\left(C_1,\dots,C_N\right)$, where $C_n$ corresponds to the distortion power for the $n$'th antenna, the instantaneous \ac{SNDR} is simplified to
\begin{align}\label{SNDRMISO}
    \gamma = \frac{P\sum_{n=1}^{N}\left|\tilde{h}_n\right|^2}{\frac{\sum_{n=1}^{N}C_n\left|\tilde{h}_n\right|^2}{\sum_{n=1}^{N}\left|\tilde{h}_n\right|^2}+\sigma^2},
\end{align}
where $\left|\tilde{h}_n\right|^2$ and $C_n$ may be calculated using Lemma \ref{BussgangLemma}. Noting that the input power to the $n$'th RX-chain is $P|h_n|^{2}$, we have
\begin{align}
    &\left|\tilde{h}_n\right|^2=\left|{h}_n\right|^2\left|\sum_{k=0}^{L-1}a_{2k+1,n}(k+1)!~P^k|h_n|^{2k}\right|^2,
\end{align}
and
\begin{align}
    C_n=\sum_{k=1}^{2L-1}\left(k!~P^k|h_n|^{2k}\sum_{i=1}^{k}a_{2i-1,n}\bar{a}_{2k-2i+1,n}\right) - \left|\tilde{h}_n\right|^2P,
\end{align}
where the variables $a_{k,n}$ indicate the coefficients for the $n$'th antenna RX-chain non-linear function in \eqref{memoryLessModel}.

For the case with per-antenna \ac{AGC}, since $\tilde{h}_n=\Tilde{g}{h}_n$ and $C_n=\kappa\left|{h}_n\right|^2P$, the SNDR in \eqref{SNDRMISO} is simplified to
\begin{align}\label{MISOSNDR}
	\gamma = \frac{P|\Tilde{g}|^2\sum_{n=1}^{N}\left|{h}_n\right|^2}{\kappa P\frac{\sum_{n=1}^{N}\left|{h}_n\right|^4}{\sum_{n=1}^{N}\left|{h}_n\right|^2}+\sigma^2}.
\end{align}

As remarked before, most of the results and conclusions of this paper can be extended to a multi-user scenario, especially if the \acp{UE} are well-separated, or when the hardware distortion is the main source of error compared to the inter-user interference. In a multi-user scenario, the \ac{SNDR} in \eqref{SNDRMISO} and \eqref{MISOSNDR} would contain an additional term in the denominator corresponding to the inter-user interference.

We are now interested in finding closed-form approximations  for the SNDR to facilitate the \ac{LIS} scaling analysis. We assume the \ac{LIS} and UE configuration as depicted in Fig.~$\ref{ArrayConfig}$. This specific \ac{LIS}-UE configuration is selected to reduce the complexity of the analytical derivations;  however, this loss of generality has reduced impact when considering general scenarios with a large enough \ac{LIS} \cite{HushaPos}.

If we assume that the \ac{LIS} consists of $\sqrt{N}\times\sqrt{N}$ square antenna elements with effective area $A$, placed edge to edge such that the distance between adjacent antennas is $\sqrt{A}$, we can approximate the summations in \eqref{MISOSNDR} by using the Riemann Integral approximation \cite{Calculus}. Since $N\gg1$ in \ac{LIS} scenarios, we can approximate the \ac{LIS} surface as a disk with radius $R$, with the same area as the square-shaped \ac{LIS}, i.e. $NA=\pi R^2$. Therefore, the $N$ antenna elements are assumed in a disk-shaped region $\mathcal{S}$ with a distance smaller than $R$ from the center of the \ac{LIS}, i.e., the antennas with $x_n^2+y_n^2<R$. Numerical results show that this approximation results in negligible error in the summation approximations when $N\gg1$. We have
\begin{align} \label{ReimannNorm2}
\sum_{n=1}^{N}|h_n|^2&\approx\iint\limits_\mathcal{S}|h(x,y)|^2dxdy \nonumber \\ 
&=\int_{\phi=0}^{2\pi}\int_{r=0}^{R}|h(r)|^2rdrd\phi,
\end{align}
where $|h(r)|^2$ corresponds to the polar representation of the channel gain, with variable changes $x=r\cos{\phi}$, $y=r\sin{\phi}$. Similarly, we can approximate the other summation appearing in \eqref{MISOSNDR} by
\begin{align}\label{IntegApprox}
    \sum_{n=1}^{N}|h_n|^4\approx A\int_{\phi=0}^{2\pi}\int_{r=0}^{R}|h(r)|^4rdrd\phi.
\end{align}
The following lemma gives close-form approximations for these summations, which can be leveraged to further develop the SNDR expression in \eqref{MISOSNDR}.
\begin{lemma} \label{Reimann1}
Assume we have a \ac{LIS} with $N\gg 1$ antenna elements with $x_n^2+y_n^2<R$, each with effective area $A$. We can then approximate the summations in \eqref{MISOSNDR} by
\begin{align} \label{Riemannform1}
 	\sum_{n=1}^{N}|h_n|^2\approx\frac{1}{2}\left(1-\frac{d}{\sqrt{d^2+R^2}}\right),
\end{align}
\begin{align} \label{Riemannform2}
 	\sum_{n=1}^{N}|h_n|^4\approx\frac{A}{32\pi}\left(\frac{1}{d^2}-\frac{d^2}{(d^2+R^2)^2}\right),
\end{align}
and their ratio as
\begin{align} \label{Riemannform3}
 	\frac{\sum_{n=1}^{N}\left|{h}_n\right|^4}{\sum_{n=1}^{N}\left|{h}_n\right|^2}\approx\frac{A}{16\pi}\left(1+\frac{d}{\sqrt{d^2+R^2}}\right)\left(\frac{1}{d^2}+\frac{1}{d^2+R^2}\right).
\end{align}\\
\begin{proof}
See Appendix \ref{ProofsAppendix}.
\end{proof}
\end{lemma}
\begin{figure*}[!t]
	\normalsize
	\begin{align} \label{SNDRonlyR}
        \gamma_\text{AGC}\approx\frac{\frac{P}{2}\left|\sum_{k=0}^{L-1}\hat{a}_{2k+1}(k+1)!\right|^2\left(1-\frac{d}{\sqrt{d^2+R^2}}\right)}{\frac{PA}{16\pi}\left[\sum_{k=1}^{2L-1}\left(k!~\sum_{i=1}^{k}\hat{a}_{2i-1}\bar{\hat{a}}_{2k-2i+1}\right) - \left|\sum_{k=0}^{L-1}\hat{a}_{2k+1}(k+1)!\right|^2\right]\left(1+\frac{d}{\sqrt{d^2+R^2}}\right)\left(\frac{1}{d^2}+\frac{1}{d^2+R^2}\right)+\sigma^2}.
	\end{align}
	\hrulefill
	\vspace*{4pt}
\end{figure*}
By adopting Lemma \ref{Reimann1} and the Bussgang parameters in \eqref{BussgangAGCg0} and \eqref{BussgangAGCDist}, we reach the SNDR approximate expression in \eqref{SNDRonlyR} at the top of this page. This SNDR approximation also applies to the additive distortion model, widely used in the literature \cite{Emil2014}.

We can also exploit the derived expressions from Lemma \ref{Reimann1} to perform an asymptotic analysis when the \ac{LIS} size grows without bounds, leading to
\begin{align} \label{halfSpace}
	&\lim_{R\rightarrow\infty}\sum_{n=1}^{N}|h_n|^2=\frac{1}{2}, \\
	&\lim_{R\rightarrow\infty}\frac{\sum_{n=1}^{N}|h_n|^4}{\sum_{n=1}^{N}|h_n|^2}=\frac{A}{16\pi d^2},
 \end{align}
 and
 \begin{align}
	&\lim_{R\rightarrow\infty}\gamma_\text{AGC}= \frac{\frac{1}{2}|\Tilde{g}|^2P}{\frac{\kappa PA}{16\pi d^2}+\sigma^2}.
\end{align}
The asymptotic channel gain in \eqref{halfSpace} implies that, for an infinitely large \ac{LIS} deployed as a plane in front on the UE, the channel gain approaches $0.5$, i.e. half of the transmitted power is received by the \ac{LIS} if there are no losses, in accordance with the law of conservation of energy since half of the space will be covered by an infinitely large \ac{LIS}. The other two asymptotic results show that, even for an infinitely large \ac{LIS}, the hardware distortion does not vanish completely. However, we shall see that the hardware distortion may be negligible in some cases, depending on hardware quality.

\subsection{The cost of ideal-hardware assumption}
In many analytical results and preliminary studies on \ac{LIS}, e.g. \cite{ShaHuBeyond1,HushaPos,Dardari2020}, ideal hardware components are assumed. When the \ac{LIS} is implemented, hardware distortion and other non-ideal effects are unavoidable. However, the performance gap can generally be covered by scaling the \ac{LIS}, which increases the array gain, allowing higher spatial multiplexing, and causes the uncorrelated hardware distortion to average out. To have a better understanding of the effects of hardware distortion when scaling the \ac{LIS}, we can study the required number of receive antennas to achieve a minimum performance requirement. In particular, we consider a fixed number of receive antennas with ideal RX-chains and find the minimum number of antennas with non-ideal RX-chains to achieve the same performance. Such analysis may be useful to better understand how much one should scale the system when adopting real-world non-ideal hardware. 

For analytical tractability, we mainly focus on the case with perfect per-antenna \ac{AGC} in this part. Let us first define the ideal-hardware case as the reference point for the analysis. If all the RX-chains are equipped with ideal hardware, i.e. $\kappa=0$, and the radius of the reception area is $R=R_0$, the SNDR becomes
\begin{align} \label{sndrRef}
	\gamma_0\approx{\frac{P}{2\sigma^2}\left(1-\frac{d}{\sqrt{d^2+R_0^2}}\right)}.
\end{align}
We are interested in finding the minimum number of RX-chains, equivalent to the minimum reception radius $R$, with non-ideal hardware components, i.e. when $\kappa >0$, to achieve the same level of SNDR as given in \eqref{sndrRef}. From \eqref{SNDRonlyR}, we would like to find the minimum $R$ fulfilling the following inequality 
\begin{align} \label{sndrDensity}
	\frac{\frac{P|\Tilde{g}|^2}{2}\left(1-\frac{d}{\sqrt{d^2+R^2}}\right)}{\frac{\kappa P A}{16\pi}\left(1+\frac{d}{\sqrt{d^2+R^2}}\right)\left(\frac{1}{d^2}+\frac{1}{d^2+R^2}\right)+\sigma^2}\geq\gamma_0.
\end{align}

Finding the minimum $R$ is in general highly non-trivial. However, low complexity algorithms can effectively approximate the solution to this inequality. An algorithm to solve this problem numerically is introduced in Appendix \ref{AlgsAppendix}.

Another approach is to define an approximate expression for \eqref{sndrDensity} and find bounds on $R$. Let us define the distortion power, which is the denominator in \eqref{sndrDensity} excluding the noise power, as
\begin{align} \label{distortionBeta}
	\beta=\frac{\kappa P A}{16\pi}\left(1+\frac{d}{\sqrt{d^2+R^2}}\right)\left(\frac{1}{d^2}+\frac{1}{d^2+R^2}\right).
\end{align}
The distortion power $\beta$ is a decreasing function of $R$, i.e. it reduces if we have more antennas for reception. The minimum and maximum of the distortion power can be found for $R\to \infty$ and $R=0$, respectively, as
\begin{align} \label{BetaMinMax}
	\beta_\text{min}= \frac{\kappa PA}{16\pi d^2}, \\
 \beta_\text{max}=\frac{\kappa PA}{4\pi d^2}. \nonumber
\end{align}

To find an upper bound on the minimum $R$ for the inequality \eqref{sndrDensity} to hold, we can consider the worst case scenario, i.e., $\beta=\beta_\text{max}$, and rewrite the inequality as
\begin{align} \label{sndrDensity2}
	{\left(1-\frac{d}{\sqrt{d^2+R^2}}\right)}&\geq\frac{\beta_\text{max}+\sigma^2}{|\Tilde{g}|^2\sigma^2}\left(1-\frac{d}{\sqrt{d^2+R_0^2}}\right) \nonumber \\
	&=\delta_\text{max}\left(1-\Upsilon\right),
\end{align}
where
\begin{align}
\delta_\text{max}=\frac{\beta_\text{max}+\sigma^2}{|\Tilde{g}|^2\sigma^2}
\end{align}
and
\begin{align}
\Upsilon=\frac{d}{\sqrt{d^2+R_0^2}}.
\end{align}
The variable $\delta_\text{max}$ is a factor containing all the imperfections, i.e., noise, gain compression, and distortion. The upper bound for the minimum $R$ can be calculated as
\begin{align} \label{Rub}
	R_\mathrm{ub}^2=d^2\left(\frac{1}{(1+\delta_\text{max}\Upsilon-\delta_\text{max})^2}-1\right),
\end{align}
which means any $R\geq R_\mathrm{ub}$ guarantees that \eqref{sndrDensity} holds.

Similarly, a lower bound on the minimum $R$ can be found by substituting $\beta_\text{max}$ for $\beta_\text{min}$ in \eqref{sndrDensity2}, where we can now define $\delta_\text{min}=\frac{\beta_\text{min}+\sigma^2}{|\Tilde{g}|^2\sigma^2}$ to find
\begin{align} \label{Rlb}
	R_\mathrm{lb}^2=d^2\left(\frac{1}{(1+\delta_\text{min}\Upsilon-\delta_\text{min})^2}-1\right).
\end{align}

An important detail to keep in mind in this analysis is the feasibility of fulfilling \eqref{sndrDensity}. As an example, imagine that the reference ideal point is selected too optimistic, i.e., $R \gg 0$; then, inequality \eqref{sndrDensity} can not hold even if $R\rightarrow\infty$. To formulate feasibility condition, we evaluate \eqref{sndrDensity} for $R\to \infty$, which results in the distortion power to be $\beta=\beta_\text{min}$. Therefore, the feasibility of finding an $R$ to satisfy \eqref{sndrDensity} is equivalent to the feasibility of finding an $R$ that satisfies
\begin{align} \label{sndrDensity3}
	{\left(1-\frac{d}{\sqrt{d^2+R^2}}\right)}&\geq\delta_\text{min}\left(1-\Upsilon\right),
\end{align}
which is only possible if $\delta_\text{min}\left(1-\Upsilon\right)\leq1$, resulting the following condition on $R_0$,
\begin{align} \label{feasR0}
	R_0^2\leq d^2\left(\frac{\delta_\text{min}^2}{(\delta_\text{min}-1)^2}-1\right).
\end{align}
The previous expression thus gives the maximum $R_0$ for which we can still guarantee the existence of an $R$ fulfilling \eqref{sndrDensity}.

\section{\ac{LIS} Antenna Selection} \label{Sec_Antenna_Selection}
As we have seen in previous sections, and verified through numerical results in Section \ref{Sec_Numerical}, hardware distortion can be a major limiting factor when scaling up \ac{LIS} systems. Introducing more back-off, employing higher quality hardware components, or using more transceiver chains can be seen as alternatives to improve system performance affected by hardware distortion \cite{Muris2021,massivemimobook}. However, all these solutions result in lower \ac{EE} and cost-efficiency of the \ac{LIS} systems, which can pose a burden in future wireless communication systems \cite{EE_new}. A potential solution is to adopt antenna selection schemes without using more resources in terms of hardware quality or extra antennas, to improve the system performance, which results in improvement of \ac{EE}.

The contribution of each antenna to the signal and distortion power depends on the received power to that individual antenna. Antennas receiving more power may contribute more to the useful signal power at the cost of introducing more distortion. In Section \ref{SNDR_Characterization}, we have seen how the receiver array size, given by the \ac{LIS} radius, affects the performance of the whole system. We assumed that the area for signal reception was selected as a disk around the center of the array, assuming that the \ac{UE} was also aligned with the \ac{CPL}. The question that arises here would be: Can we improve the signal reception performance, i.e. SNDR, by selecting the antennas from another region? In other words, can we get an improvement from considering a reception area with a different shape than a disk with a center aligned with the UE position? This question can be formulated as an antenna selection problem, where we consider a \ac{LIS} with $N\gg 1$ elements, and a resource constraint which forces the LIS to only use $N_\text{max}$ of the antenna elements for signal reception.

If the \ac{LIS} is restricted to perform MRC on the received signals from the selected antennas, the optimization problem for antenna selection can be formulated as
\begin{align} \label{generalOpt1}
	\max_{z_1, \dots, z_N } ~ &\frac{P\sum_{n=1}^{N}z_n\left|\tilde{h}_n\right|^2}{\frac{\sum_{n=1}^{N}z_nC_n\left|\tilde{h}_n\right|^2}{\sum_{n=1}^{N}z_n\left|\tilde{h}_n\right|^2}+\sigma^2},\\
	&\mathrm{s.t.}~~ z_n\in \{0,1\} ~~~~ \forall n \in \{1, \dots, N \}\nonumber \\
	&~~~~~ \sum_{n=1}^{N}z_n\leq N_\text{max} \nonumber ,
\end{align}
where $z_n$ is the binary antenna selection parameter. In general, this problem cannot be solved in closed-form. However, the \ac{LIS} can perform a simple low-complexity heuristic search to find the optimal solution.
    
For the case with perfect \ac{AGC}, we can simplify the optimization problem \eqref{generalOpt1} by exploiting the \ac{SNDR} approximations from Section \ref{SNDR_Characterization}. Let us consider the setup from Section \ref{SystemModel} to simplify this problem. Since the transmitter is located at $(0,0,d)$, i.e. on the boresight of the \ac{LIS}, the antennas with the same distance from the center of the array have equal received power. Therefore, if an antenna is in the set of the selected antennas, all the antennas with the same distance from the center should also be in that set unless the maximum number of selected received antennas is reached. This implies that the optimal solution for the selected antennas is in general an annulus, i.e., the region between two concentric circles.

Let us define the set of selected antennas as $\mathcal{F}=\{i\in\mathbb{N}~\vert~r<x_i^2+y_i^2<R\}$. Therefore, we have $z_i=1$ if $i \in \mathcal{F}$. By leveraging Lemma~\ref{Reimann1}, the following approximation then holds for the numerator in \eqref{generalOpt1},
\begin{align}	P\sum_{n=1}^{N}z_n\left|\tilde{h}_n\right|^2&=P\left|\Tilde{g}\right|^2\sum_{n\in\mathcal{F}}|h_n|^2 \\
        &\approx\frac{Pd\left|\Tilde{g}\right|^2}{2}\left(\frac{1}{\sqrt{d^2+r^2}}-\frac{1}{\sqrt{d^2+R^2}}\right) \nonumber,
\end{align}
and for the denominator, we have
\begin{align}
	&\frac{\sum_{n=1}^{N}z_nC_n\left|\tilde{h}_n\right|^2}{\sum_{n=1}^{N}z_n\left|\tilde{h}_n\right|^2}=\frac{P\kappa\sum_{n\in\mathcal{F}}\left|{h}_n\right|^4}{\sum_{n\in\mathcal{F}}|h_n|^2} \\
	&\approx\frac{P\kappa dA}{16\pi}\left(\frac{1}{\sqrt{d^2+r^2}}+\frac{1}{\sqrt{d^2+R^2}}\right)\left(\frac{1}{d^2+r^2}+\frac{1}{d^2+R^2}\right) \nonumber,
\end{align}
where $\Tilde{g}$ and $\kappa$ are given in \eqref{BussgangAGCg0} and \eqref{BussgangAGCDist}. By exploiting the above approximations, the optimization problem \eqref{generalOpt1} converts to
\begin{align} \label{rRoptprob1}
		\max_{r,R} ~ &\frac{\frac{Pd\left|\Tilde{g}\right|^2}{2}\left(\frac{1}{\sqrt{d^2+r^2}}-\frac{1}{\sqrt{d^2+R^2}}\right)}{\frac{P\kappa dA}{16\pi}\left(\frac{1}{\sqrt{d^2+r^2}}+\frac{1}{\sqrt{d^2+R^2}}\right)\left(\frac{1}{d^2+r^2}+\frac{1}{d^2+R^2}\right)+\sigma^2},\\
		&\mathrm{s.t.}~~ r,R \geq 0 \nonumber \\
		&~~~~~ \pi(R^2-r^2)\leq N_{max}A \nonumber
\end{align}
which has the benefit of being a continuous function of only two variables, and it has a reduced complexity over \eqref{generalOpt1} when solved by numerical methods. An algorithm to solve this problem numerically is introduced in Appendix \ref{AlgsAppendix}. Numerical results in Section \ref{Sec_Numerical} show that solving the approximated optimization problem \eqref{rRoptprob1} results in negligible performance loss compared to directly solving \eqref{generalOpt1}.

\section{\ac{LIS} Panel Selection} \label{Sec_Panel_Selection}
The analysis so far was based on the assumption that \acp{LIS} are going to be implemented as very large surfaces with $N\gg 1$ antennas equally spaced on a plane, e.g., a wall or a ceiling of a building. As mentioned earlier, this is a common vision for \ac{LIS} implementations in future wireless networks. However, it may be impractical when considering current base station technologies, e.g., massive \ac{MIMO}. A more favorable option, without losing practicality and cost-efficiency, is to implement \ac{LIS} as a grid of $N_\mathrm{p}$ distributed panels, each equipped with $M$ uniformly spaced antennas. Therefore, we extend our previous results to a panel-based \ac{LIS} scenario.

Let us assume that $M\ll N_\mathrm{p}$ and the UE is far enough to be the far-field for each panel such that the channel gain amplitude $|h_n|^2$ has a negligible variation for all the antenna elements on each panel, and the only significant change across the panel is carried in the phase. Therefore, the \ac{SNDR} after applying MRC is given by
\begin{align} \label{sndrPanels}
	\gamma=\frac{P M \sum_{n=1}^{N_\mathrm{p}}|\tilde{h}_n|^2}{\frac{\sum_{n=1}^{N_\mathrm{p}}C_n|\tilde{h}_n|^2}{\sum_{n=1}^{N_\mathrm{p}}|\tilde{h}_n|^2}+\sigma^2},
\end{align}
where $|\tilde{h}_n|$ and $C_n$ correspond to the channel gain and the distortion power for all the antennas from the $n$'th panel. By comparing \eqref{sndrPanels} and \eqref{SNDRMISO}, we can see that they look the same except for the panel array-gain factor $M$. However, we should note that $|\tilde{h}_n|$ has a slightly different meaning now since it corresponds to the common channel gain of the whole panel instead of the channel gain for each antenna. Nevertheless, we expect a similar behavior in performance when scaling this type of \ac{LIS}.

Taking into account \eqref{sndrPanels}, we can transform the antenna selection problem in \eqref{generalOpt1} into a panel selection problem
\begin{align} \label{PanelOptProb}
	\max_{z_n} ~ &\frac{PM\sum_{n=1}^{N_p}z_n\left|\tilde{h}_n\right|^2}{\frac{\sum_{n=1}^{N_p}z_nC_n\left|\tilde{h}_n\right|^2}{\sum_{n=1}^{N_p}z_n\left|\tilde{h}_n\right|^2}+\sigma^2},\\
	&\mathrm{s.t.}~~ z_n\in \{0,1\} ~~~~~ n\in \{1,...,N\}\nonumber \\
	&~~~~~ \sum_{n=1}^{N_p}z_n\leq N_\text{max} \nonumber
\end{align}
where the binary variable $z_n$ determines if the $n$'th panel is active or not, and $N_\text{max}\leq N_\mathrm{p}$ is the maximum number of panels that can be selected for signal decoding. This panel selection problem can also be interpreted as the original antenna selection problem after grouping the antenna elements on the \ac{LIS} into a grid of rectangular panels.

While performing a heuristic search to solve \eqref{PanelOptProb} has reasonable complexity, it may still be beneficial, from a practical point of view, to consider closed-form sub-optimal solutions for specific cases. Such closed-form solutions can also provide some intuition about how to optimize the design of \ac{LIS} architectures, e.g., for defining suitable panel placement strategies. For analytical tractability, we will focus on the 3rd order non-linear model, i.e., where the hardware model is given by \eqref{memoryLessModel}, with $L=2$, which retains much of the practicality since the 3rd order non-linearity is known to be the dominant source of in-band hardware distortion \cite{DemirChannel,schenk2008rf}. 
Nevertheless, in case higher orders of non-linearity are of interest, one can fit the higher order model to a 3rd order model and adopt our proposed solution with the cost of a marginal deviation from the optimum.

Let us start with the case where only one panel is selected, i.e., $N_\text{max}=1$. In this case, the panel selection problem boils down to optimizing the input power for a SISO scenario. The reason is that we can find the optimum input power for the SISO scenario and select the panel with the closest input power to that optimum value. In the general case, where $N_{\max} \geq 1$, the \ac{LIS} can simply select the $N_\text{max}$ panels with the input power closest to the optimum value. Therefore, finding the optimum input power for the SISO case is of great importance.
For a SISO scenario where a symbol $s \sim \mathcal{CN}(0,P)$ is transmitted over the channel ${h} \in \mathbb{C}$, the input to the RX-chain will be $x\sim \mathcal{CN}(0,\rho)$, where $\rho=P|h|^2$. If the receiver RX-chain has a 3rd order non-linearity as described above, from \eqref{3rdg} and \eqref{3rdGKappa}, the \ac{SNDR} may be expressed as
\begin{align}
	\gamma_\text{SISO} = \frac{|a_1+2a_3\rho|^2\rho}{2|a_3|^2\rho^3+\sigma^2}.
\end{align}

As previously motivated, we are interested in solving the optimization problem
\begin{align} \label{SISOOpt}
	\max_{\rho} ~ &\frac{|a_1+2a_3\rho|^2\rho}{2|a_3|^2\rho^3+\sigma^2},\\
	&\mathrm{s.t.}~~  0<\rho<\rho_{\max} \nonumber&
\end{align}
where $\rho_{\max}$ is the maximum expected received power, which is imposed by the specific scenario. Note that the optimization variable $\rho$ can be altered either by adjusting $P$, e.g., via power control, or $|h|^2$, e.g., via UE movements or panel selection. Since we are interested in this problem only for the purpose of panel selection, the transmit power is assumed to be fixed, while $|h|^2$ can be controlled by selecting different panels across the \ac{LIS}.

The optimization problem \eqref{SISOOpt} is concave and finding its optimal solution in the current format, involves solving a 5th order equation after taking the first derivative. However, we can try to find a suitable approximation for the numerator and simplify the objective function. For example, we can approximate the term $|a_1+2a_3\rho|^2$ in the numerator as a linear function of $\rho$, $\alpha+\beta\rho$, for $0<\rho<\rho_{\max}$. According to \eqref{coeffsNorm}, we have $a_1=\hat{a}_1$ and $a_3=\hat{a}_3/\rho_{\max}$, which gives approximation parameters $\alpha = |\hat{a}_1|^2$ and
\begin{align} 
	\beta = \frac{|\hat{a}_1+2\hat{a}_3|^2-|\hat{a}_1|^2}{\rho_{\max}}.
\end{align}
Applying the above approximation and taking the first derivative of the approximate \ac{SNDR}, we end up with the 4th order equation
\begin{align} \label{4thorder}
	2\beta|{a}_3|^2\rho^4+4\alpha|{a}_3|^2\rho^3-2\beta\sigma^2\rho-\alpha\sigma^2=0.
\end{align}
Considering amplifier characterizations from \cite{Ericsson2016}, we note that the factor $2\beta|\hat{a}_3|^2$ is much smaller than other factors in the equation. Therefore, we approximate \eqref{4thorder} by
\begin{align}
	\rho^3-\frac{\beta\sigma^2}{2\alpha|{a}_3|^2}\rho-\frac{\sigma^2}{4|{a}_3|^2}=0,
\end{align}
which is a depressed cubic equation $\rho^3+c_1\rho+c_0=0$ with
\begin{align}
	c_0=-\frac{\sigma^2}{4|{a}_3|^2}
\end{align}
and
\begin{align}
	c_1=-\frac{\beta\sigma^2}{2\alpha|{a}_3|^2}.
\end{align}
According to Cardano's formula \cite{cardano}, and since $\Delta=c_0^2/4+c_1^3/27$ is positive, this equation has only one real solution given by
\begin{align} \label{closedFrom3rd}
	\rho_\text{opt} = \sqrt[3]{-\frac{c_0}{2}+\sqrt\Delta} + \sqrt[3]{-\frac{c_0}{2}-\sqrt\Delta}.
\end{align}
The numerical results in Section \ref{Sec_Numerical} indicate a high accuracy in the approximation. Hence, \eqref{closedFrom3rd} provides a suitable closed-form approximate solution to the panel selection problem.

\section{Numerical Results} \label{Sec_Numerical}
In this section, we provide numerical examples to gain more insights about the aforementioned methods and derivations. As performance metric, we consider a lower bound on the spectral efficiency given by $R_\text{a}=\text{log}_2(1+\gamma)$, which corresponds to an achievable rate. This achievable rate comes from assuming a Gaussian additive distortion term in the Bussgang decomposition, corresponding to the worst case scenario \cite{worst_noise}. The UE is assumed to be at distance $d=25\lambda$ from the center of \ac{LIS} transmitting with transmit power $P$ such that SNR$=10dB$ at the center of \ac{LIS}. The gray areas in all the figures in this section illustrate the range of parameters for which the UE is in far-field of the \ac{LIS}. For the hardware distortion model, we use the measurements from one of the 3GPP reports \cite{Ericsson2016}, which also estimates the parameters for the memory-less polynomial model \eqref{memoryLessModel} based on real-world RF hardware at different frequencies and bandwidths. To have a better understanding of the difference in hardware distortion effects for different RF components, we have employed \eqref{BussgangAGCg0} and \eqref{BussgangAGCDist} to calculate $\Tilde{g}$ and $\kappa$ based on a range of measurements in \cite{Ericsson2016}, which have been performed at the frequencies $2.1$ GHz and $28$ GHz, for GaA, CMOS, and GaN amplifiers. Table \ref{KappaG0} summarizes the results. Note that, for ideal hardware, we would have $\vert \Tilde{g} \vert^2=1$ and $\kappa = 0$.
\begin{table}[h!]
\centering
\begin{tabular}{ |p{0.8cm}||p{1cm}|p{1.10cm}|p{0.75cm}|p{0.7cm}|p{0.7cm}|  }
 \hline
 Type&Freq.&BW&BO&$\kappa$&$|\Tilde{g}|^2$\\
 \hline
 GaA&   $2.1$ GHz  & $40$ MHz   &$10$ dB & $0.208$&$0.937$\\
 CMOS   & $28$ GHz    &$400$ MHz&   $10$ dB& $0.252$&$0.894$\\
 GaN &$2.1$ GHz & $40$ MHz&  $8$ dB& $0.035$&$0.811$\\
 GaN    &$28$ GHz & $400$ MHz&  $8$ dB& $0.132$&$0.884$\\
 \hline
\end{tabular}
\caption{Examples of Hardware Distortion Parameters in \eqref{BussgangAGCg0} and \eqref{BussgangAGCDist} for  different amplifier types at specific frequencies, bandwidth(BW), and back-off (BO).\\}
\label{KappaG0}
\end{table}
\begin{figure}[t]
	\centering
	\includegraphics[width=3.2in]{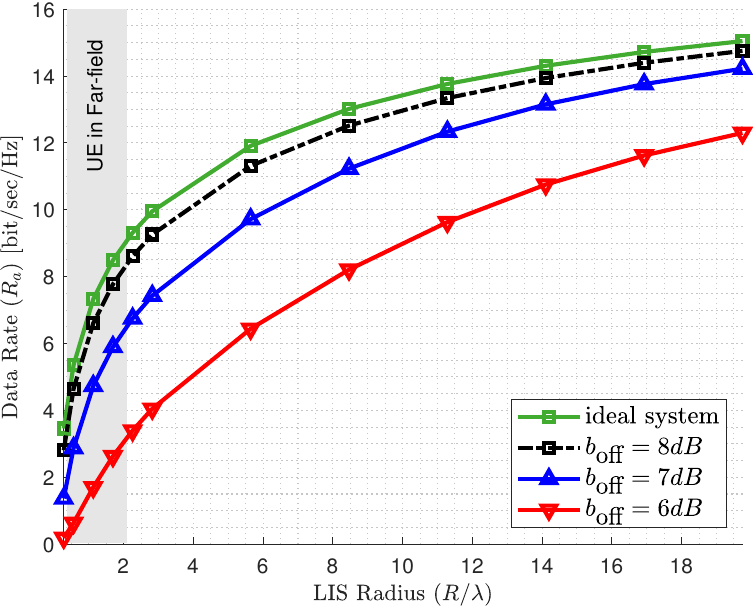}
	\caption{Achievable data rate vs \ac{LIS} radius in terms of $\lambda$. The UE is at distance $d=25\lambda$ from the center of \ac{LIS}.}
	\label{SE_vs_R}
\end{figure}

\subsection{Scaling Analysis Example}
We begin with studying the scaling performance of \acp{LIS} for different levels of hardware distortion. Let us consider the model based on a Gallium Nitride (GaN) amplifier designed for operation at 2.1 GHz as a case study, where the model parameters have been estimated from input-output measurements at a sample rate of 200 MHz and a signal bandwidth of 40 MHz. We would have similar results and conclusions if we select another data set from \cite{Ericsson2016} and table \ref{KappaG0}. The report provides the coefficients for normalized inputs, i.e., $\hat{a}_{2k+1}$ in \eqref{coeffsNorm} and \eqref{coeffsNormAGC}, for 3rd, 5th, 7th, and 9th order non-linearity. We consider a discrete \ac{LIS} with antenna elements separated by $\lambda/2$ on a rectangular grid. Selecting this value for antenna-spacing is based on the fact that $\lambda/2$ spacing allows capturing the full spatial degrees of freedom, while facilitating practical implementation \cite{ShaHuBeyond1,LIS_sampled}. Nevertheless, it can be verified that under the considered framework, the same conclusions can be derived for antenna spacings below $\lambda/2$.

In Fig.~\ref{SE_vs_R}, we illustrate the achievable rate $R_\text{a}$ for different levels of back-off in \eqref{coeffsNorm}. By comparing our results to the performance of an ideal system, we can see that the hardware distortion can degrade the system performance significantly even if we use $7$ dB of back-off, which is a typical value for low-power receivers. One should also note that for a sufficiently large back-off, e.g. $8$ dB or higher, the performance approaches the ideal case at the cost of low energy efficiency in the RF amplifiers, which is not favorable, especially in \ac{LIS} scenarios. Since the case with $8$ dB has a similar performance to the ideal case, we only focus on the lower back-off values from now on.

\begin{figure}[t]
	\centering
	\includegraphics[width=3.2in]{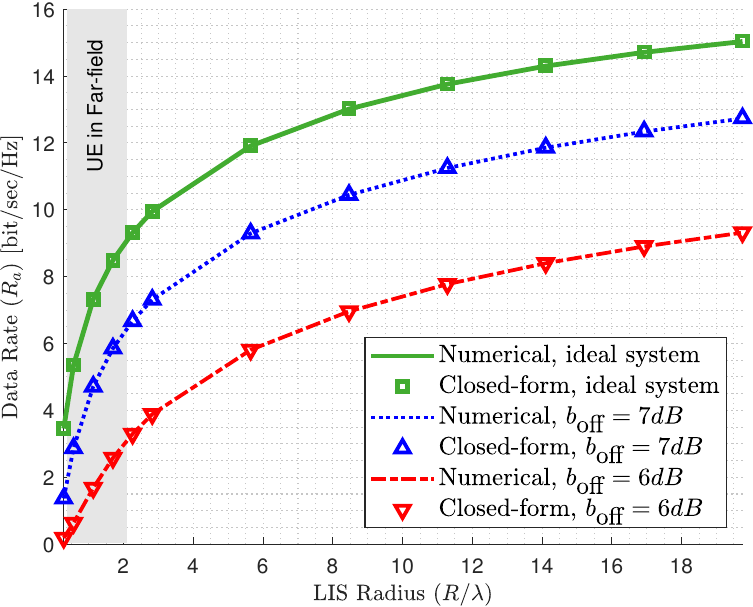}
	\caption{Numerical and Closed-form achievable data rate with perfect \ac{AGC} vs \ac{LIS} radius in terms of $\lambda$. The UE is at distance $d=25\lambda$ from the center of \ac{LIS}.}
	\label{SE_vs_R_AGC}
\end{figure}

In Fig.~\ref{SE_vs_R_AGC} we consider the case with perfect per-antenna \ac{AGC}, where we plot the achievable rate for different levels of back-off. So as to evaluate the approximation error for the proposed close-form expressions, we consider both the closed-form expression and the exact numerical values. Firstly, we can see that the close-form approximations from lemma \ref{Reimann1} overlap closely with the numerical exact values, which means the approximations are valid and can be used for analytical results with very low approximation error. As a result, we can approximate problem \eqref{generalOpt1} with \eqref{rRoptprob1} with a negligible loss in performance. Secondly, we can see that similar to Fig.~\ref{SE_vs_R}, which was without per-antenna perfect \ac{AGC}, the hardware distortion degrades the system performance significantly, which implies that hardware distortion may potentially limit the performance in \ac{LIS} scenarios no matter if we have perfect \ac{AGC} or not.

\subsection{Antenna and Panel selection}
\begin{figure}[t]
	\centering
	\includegraphics[width=3.4in]{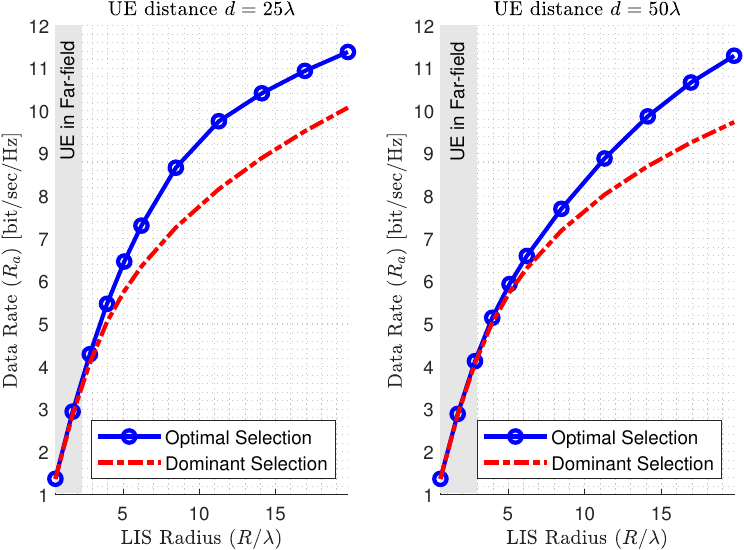}
	\caption{Achievable data rate with perfect \ac{AGC} vs \ac{LIS} radius in terms of $\lambda$ with fixed back-off for optimal antenna selection solved from \eqref{generalOpt1} and dominant antenna selection where the antennas with highest received power are selected.}
	\label{AS_SE_vs_R}
\end{figure}

In Fig.~\ref{AS_SE_vs_R}, the performance of using the proposed optimal antenna selection in comparison to using the dominant antennas, i.e., antennas with highest received power, is illustrated. Note that, in the single-\ac{UE} scenario without hardware distortion, dominant antenna selection is the optimal selection approach in terms of capacity, while for multi-\ac{UE} scenarios, the loss with respect to optimal sum-rate maximization is marginal for large enough number of RX-chains \cite{Gao2015}. We have used the 9th order distortion model for a GaN amplifier at $2.1$ GHz, as described in the previous section, with a $7$ dB of back-off. The UE transmit power is again selected such that the SNR at the center of \ac{LIS} reaches $10$ dB, which corresponds to the respective value of $P_\text{max}$ in \eqref{coeffsNorm}. We also assume that $N_\text{max}=\ceil{0.1N}$ is selected as the antenna selection constraint. As we can see in Fig. \ref{AS_SE_vs_R}, adopting the optimal antenna selection can improve the system performance significantly for medium to large \ac{LIS} radius. We can also see that the gain from adopting antenna selection is negligible if the UE is in the far-field of the \ac{LIS}.

\begin{figure}[t]
	\centering
	\includegraphics[width=3.2in]{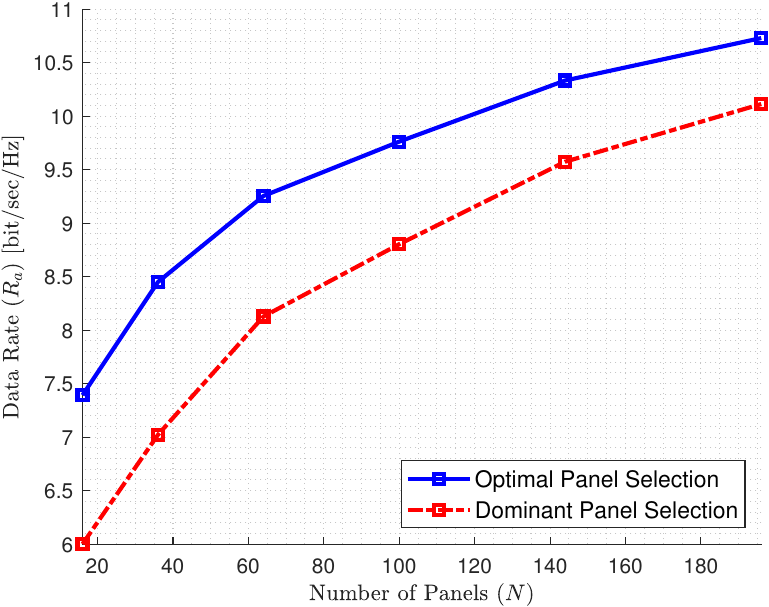}
	\caption{Achievable data rate with perfect \ac{AGC} vs number of panels with fixed back-off for optimal panel selection solved from \eqref{PanelOptProb} and dominant panel selection. The UE is at distance $d=25\lambda$ from the center of \ac{LIS}, the panel selection constraint in \eqref{PanelOptProb} is $N_\text{max}=\ceil{0.1N}$, and Each panel consists of $M=16$ antennas with $\lambda/2$ spacing. The distance between the centers of adjacent panels is $\delta_p=5\lambda$.}
	\label{PS_SE_vs_R}
\end{figure}

In Fig. \ref{PS_SE_vs_R} we have illustrated the performance of panel-based \ac{LIS} when adopting the proposed optimal panel selection versus the baseline approach corresponding to performing dominant panel selection, for different number of panels. The panel selection constraint in \eqref{PanelOptProb} is set to $N_\text{max}=\ceil{0.1N}$. We also assume that each panel consists of $M=16$ antennas with $\lambda/2$ spacing, and the distance between the centers of adjacent panels is set to $\delta_p=5\lambda$. The same distortion model as in Fig.~\ref{AS_SE_vs_R} is used for each antenna element of the \ac{LIS} panels. We can see that there is a significant gain from applying the proposed optimal panel selection. In comparison to the results from Fig.~\ref{AS_SE_vs_R} for antenna selection, the achievable performance gains are much higher, which implies that in the practical case of panel-based \ac{LIS} deployment, it is even more important to consider the panel selection schemes in the presence of hardware distortion.

\begin{figure}[t]
	\centering
	\includegraphics[width=3.2in]{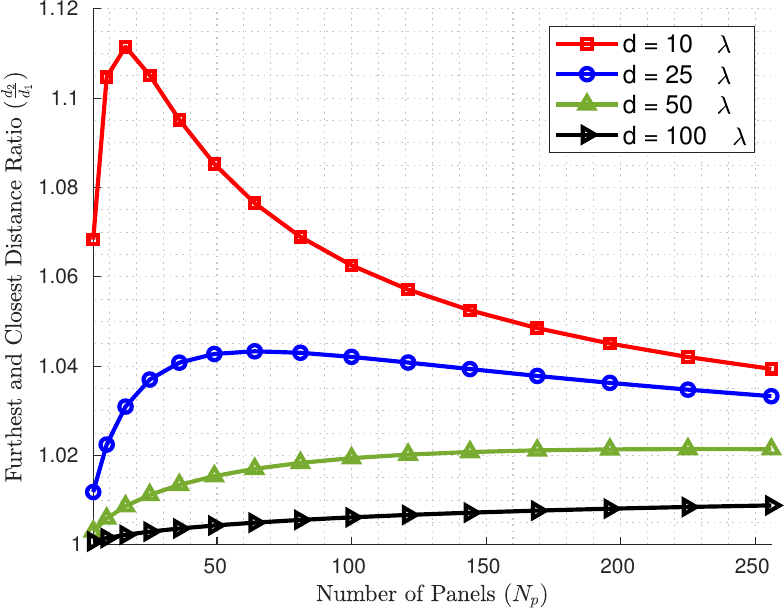}
	\caption{The maximum ratio between the furthest and closest antenna element on each panel to the UE vs the number of panels, for different distances of the \ac{UE} to the center of \ac{LIS}.}
	\label{panels_NF}
\end{figure}

{For panel-based \ac{LIS} and as previously mentioned, we have assumed that the UE is in the far-field of each panel such that we can simplify the panel selection problem into \eqref{sndrPanels}. To illustrate the validity of this assumption in this section, we need to compare the distance between the closest and furthest antenna elements on each panel to the UE, which are denoted by $d_1$ and $d_2$ respectively. In Fig. \ref{panels_NF}, we have compared the maximum ratio of these distances for the panel-based \ac{LIS} scenario as described above. We can see that the far-field assumption for individual panels holds for $d=25 \lambda$ and beyond, which is the case we have considered in this section.}

\begin{figure}[t]
	\centering
	\includegraphics[width=3.3in]{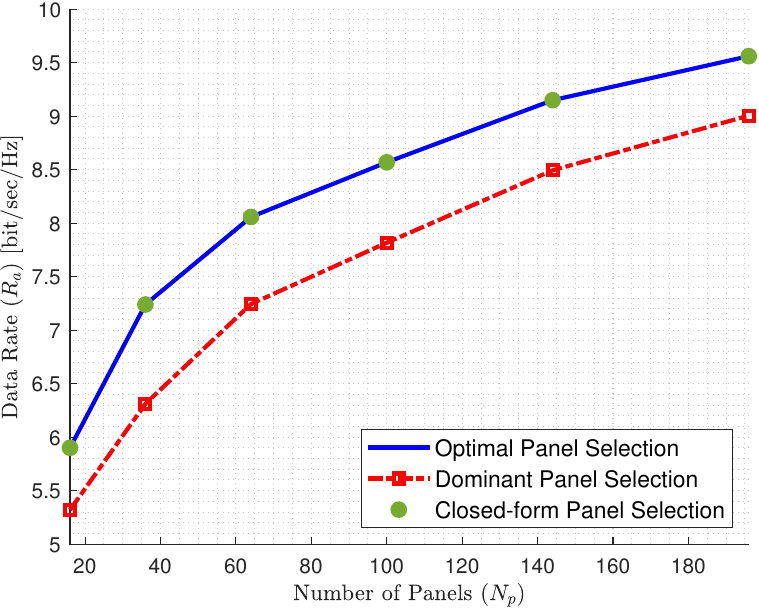}
	\caption{Achievable data rate vs number of panels with fixed back-off for optimal panel selection solved from \eqref{PanelOptProb}, close-form panel selection in \eqref{closedFrom3rd}, and dominant panel selection. The setup parameters are the same as Fig. \ref{PS_SE_vs_R}.} 
	\label{PS_CL_SE_vs_R}
\end{figure}

\begin{figure}[t]
	\centering
	\includegraphics[width=3.4in]{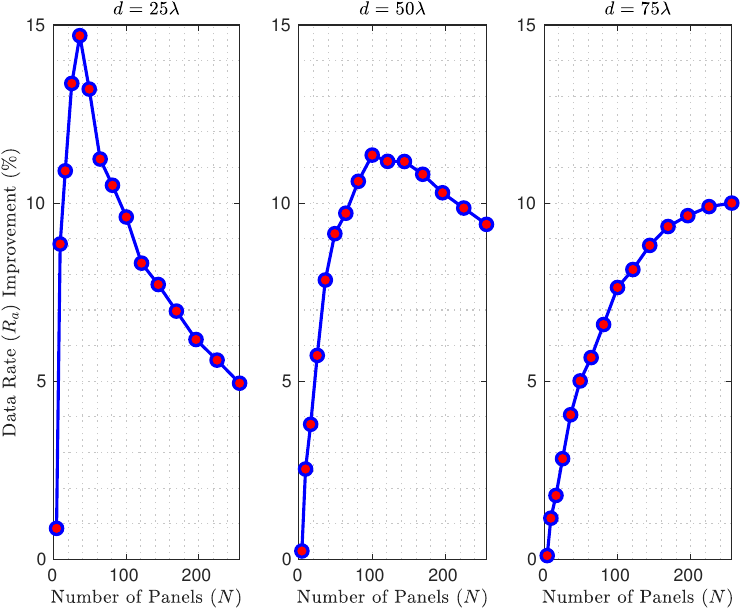}
	\caption{Percentage achievable data rate improvement for different UE distances with fixed back-off for optimal panel selection and dominant panel selection. The setup parameters are the same as Fig. \ref{PS_SE_vs_R}.}
	\label{PS_3Dist}
\end{figure}

In Fig.~\ref{PS_CL_SE_vs_R}, we have evaluated the performance of the approximate closed-form solution to the panel selection problem in comparison to the numerical optimal solution and the dominant panel selection. We have used the 3rd order non-linearity model for the hardware distortion as motivated in section \ref{Sec_Panel_Selection}. The panel selection constraint and panel placement is the same as in Fig.~\ref{PS_SE_vs_R}. As we see, the optimal panel selection and closed-form panel selection outperform the dominant panel selection significantly, and the close-form panel selection performance is very close to the numerical optimal method, which implies that the proposed close-form solution is accurate enough to use in \ac{LIS} panel selection problems with hardware distortion. In Fig.~\ref{PS_3Dist}, we compare the performance gain of adopting such schemes for 3 different distance of the UE. We can see that the gain is higher if the UE is more into the near-field of the \ac{LIS}, and it is still significant if the UE is placed at further distances. The results from Fig. 5, 6, 8, and 9 show that we can improve the system performance by performing the proposed antenna and panel selection without using more resources in terms of extra antennas and panels. This implies that the proposed selection methods can improve the system \ac{EE} which is favorable when scaling up \ac{LIS} systems for future generations of wireless networks.

\begin{figure}[t]
	\centering
	\includegraphics[width=3.3in]{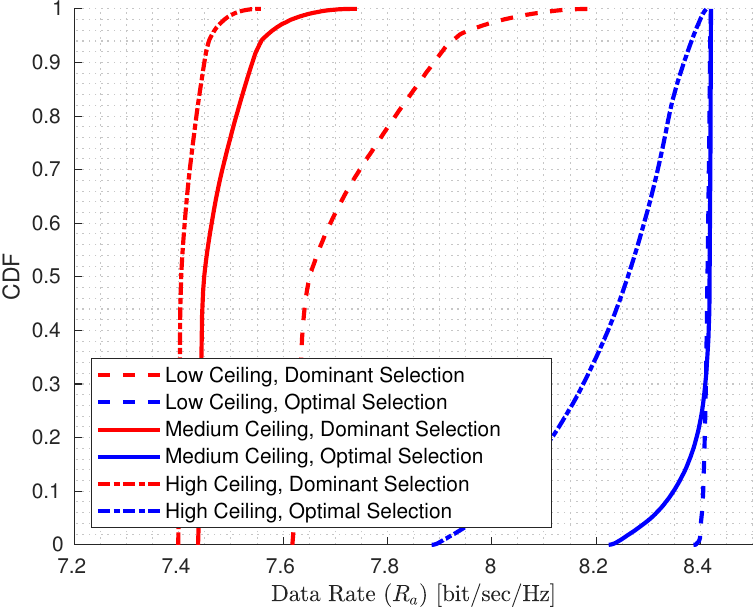}
	\caption{CDF of achievable data rate with fixed back-off for optimal panel selection and dominant panel selection for randomly located UE in a $7m \times 7m$ room with different ceiling heights. Low, medium, and high ceiling, corresponding to perpendicular distance of $25\lambda$, $50\lambda$, and $75\lambda$, between the \ac{UE} and \ac{LIS}, respectively. Other setup parameters are the same as Fig. \ref{PS_SE_vs_R}.}
	\label{CDF_Ceiling_Heights}
\end{figure}

To have a better understanding of the importance of panel selection in \ac{LIS} applications, we simulate a practical panel-based \ac{LIS} deployment scenario where a panel-based \ac{LIS} consisting of $4 \times 4$ panels covers the ceiling of a $7m \times 7m$ room. The carrier frequency is set to $2.1$ GHz to match the hardware distortion measurements for the GaN amplifier from \cite{Ericsson2016}. The distance between panels is $5\lambda$, the antenna spacing is $\lambda/2$, and the panel selection constraint is set to $N_\text{max}=\ceil{0.1N}$. With these setup parameters, there are in total $N=81$ panels covering the ceiling and the selection schemes allocate $N_\text{max}=9$ panels to serve the UE, which is randomly located at the bottom of the room. We consider three different room heights, $3.57m$, $7.14m$ and $10.71m$, corresponding to perpendicular distance of $25\lambda$, $50\lambda$, and $75\lambda$, between the \ac{UE} and \ac{LIS}, respectively. In Fig.~\ref{CDF_Ceiling_Heights}, we compare the CDF of the achievable rate for the case where the proposed optimal panel selection is adopted in comparison to the case where the dominant panels are assigned to the UE. We can see that for this scenario with typical setup parameters of a \ac{LIS} indoor application, the gain from adopting the proposed panel selection is significant, and the \ac{LIS} can provide a higher data rate with high probability just by adopting our proposed panel selection. We can also see that the data rate becomes worse in buildings with higher ceiling if the dominant panel selection is used, while the proposed panel selection can make the data rate more stable and higher, no matter if the ceiling is high or low.

\subsection{Distortion Correlation}\label{DistortionCorrelation}
As remarked in Section \ref{SNDR_Characterization}, the distortion covariance matrix $\bm{C_{\eta\eta}}$ is in general non-diagonal, and assuming it diagonal, i.e., disregarding the correlation, results in an over-estimation of the \ac{SNDR}. In this section, we study the impact of disregarding the distortion correlation on the scaling behavior and on the performance of the panel selection schemes. Assuming a 3rd order non-linearity model, i.e. $L=2$, for the non-linearity model in \eqref{memoryLessModel}, the non-diagonal distortion covariance matrix elements are given by
\begin{align}\label{Cetaeta_corr}
	[\bm{C_{\eta\eta}}]_{ij} =2a_{3,i}\bar{a}_{3,j}|\rho_{ij}|^2\rho_{ij},
\end{align}
where $\rho_{ij} = \mathbb{E}\{x_i\bar{x}_j\}$, and $x_i$ is the input signal to the $i$'th antenna. While characterizing the non-diagonal elements of $\bm{C_{\eta\eta}}$ for higher order non-linearities is beyond the scope of this paper, we expect the 3rd order case to capture the main impact of disregarding distortion correlation.

\begin{figure}[t]
    \centering
    \includegraphics[width=3.3in]{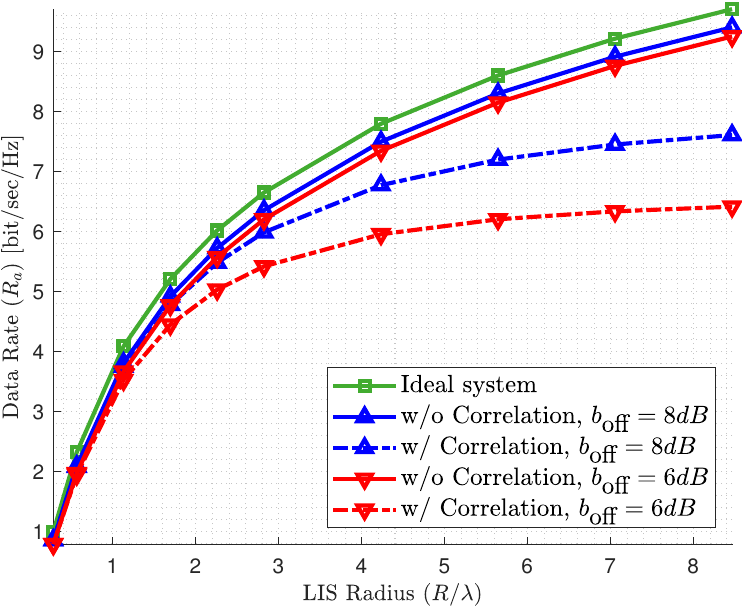}
    \caption{Achievable data rate vs \ac{LIS} radius in terms of $\lambda$, with and without considering distortion correlation. The UE is at distance $d=25\lambda$ from the center of \ac{LIS}.} 
    \label{Scaling_Correlation}
\end{figure}

In Fig.~\ref{Scaling_Correlation}, we have analyzed the scaling behavior of \ac{LIS} by comparing the achievable rate $R_\text{a}$ for different levels of back-off, both with and without considering the distortion correlation. The transmit power is selected such that $\text{SNR}=0\text{ dB}$ at the center of the \ac{LIS}. We can see that as we scale up the \ac{LIS}, the impact of disregarding the distortion grows progressively. However, for reasonably large \acp{LIS}, the impact is still quite limited. On the other hand, taking into account the distortion correlation will only strengthen the important finding that hardware distortion has a significant impact on the performance of scaled-up \ac{LIS} systems.

\begin{figure}[t]
    \centering
    \includegraphics[width=3.4in]{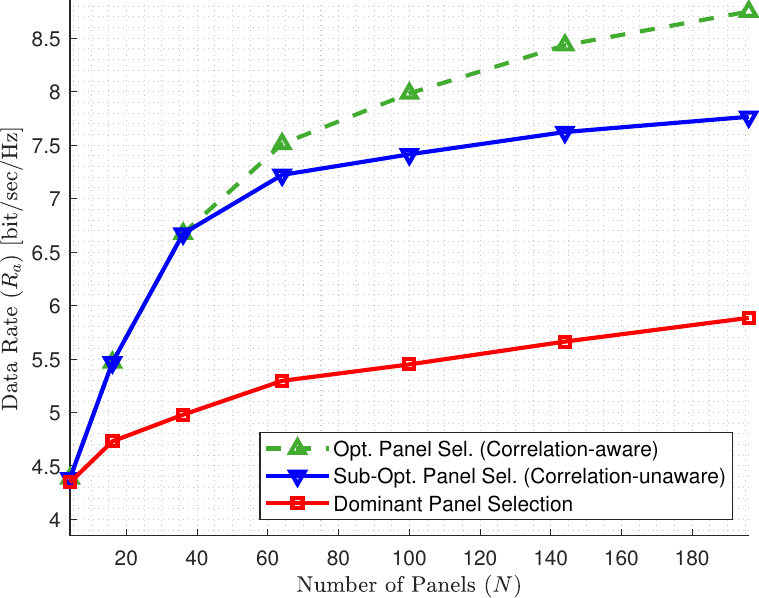}
    \caption{Achievable data rate with distortion correlation vs number of panels. The UE is at distance $d=25\lambda$ from the center of \ac{LIS}, the panel selection constraint in \eqref{PanelOptProb} is $N_\text{max}=\ceil{0.1N}$, and Each panel consists of $M=16$ antennas with $\lambda/2$ spacing. The distance between the centers of adjacent panels is $\delta_p=5\lambda$.}
    \label{PS_Correlation}
\end{figure}

In Fig.~\ref{PS_Correlation} we analyze the performance of the panel selection schemes when these are either aware or unaware of the distortion correlation. For that purpose, we have plotted the achievable data rate with hardware distortion correlation for three cases: dominant panel selection, sub-optimal panel selection which performs the panel selection unaware of the distortion correlation, and the optimal panel selection which takes into account the distortion correlation for panel selection. All the simulation parameters are the same as in Fig.~\ref{PS_CL_SE_vs_R}. We can see that, although Fig. \ref{Scaling_Correlation} shows that disregarding distortion correlation can result in considerable over-estimation of the data rate, this approximation can be leveraged for low-complexity panel selection, still resulting in close-to-optimum performance and significant gain compared to the baseline approach.

\section{Conclusion} \label{Sec_Conclusion}
In this paper, we have studied the impact of hardware distortion when considering \ac{LIS} implementations with non-ideal RX-chains. We have derived analytical expressions for the \ac{SNDR} considering the memory-less polynomial model for non-ideal hardware at the RX-chains. Such expressions enabled us to evaluate the performance of the \ac{LIS} with hardware distortion when scaling up the system. We observe that the hardware quality can effectively limit the system performance even for extremely large \acp{LIS}. We have also proposed antenna selection schemes for \ac{LIS} and we have shown that adopting such schemes can improve the performance significantly. We also consider a more practical case where the \ac{LIS} is deployed as a grid of multi-antenna panels, and define panel selection problems to improve the system performance. For the special case of 3rd-order nonlinear model, we have introduced a close-form panels selection solution, which can be exploited to efficiently switch panels with near-optimum performance.

{\appendices
\section{Proof of Lemmas}\label{ProofsAppendix}
\subsubsection{Proof of Lemma \ref{BussgangLemma}}
Considering the memory-less polynomial model \eqref{memoryLessModel} with $x_\text{in}\sim\mathcal{CN}(0,P)$, we first calculate $\mathbb{E}\left[y_\text{out}\bar{x}_\text{in}\right]$ as
\begin{align} \label{EyxProof}
    \mathbb{E}\left[y_\text{out}\bar{x}_\text{in}\right] &= \mathbb{E}\left[\sum_{k=0}^{L-1} a_{2k+1}x_\text{in}|x_\text{in}|^{2k}\bar{x}\right] \nonumber \\
    &= \sum_{k=0}^{L-1}a_{2k+1}\mathbb{E}\left[|x_\text{in}|^{2(k+1)}\right],
\end{align}
where the term $|x_\text{in}|^2$ is a Rayleigh variable, with $\mathbb{E}|x_\text{in}|^2 = P$. The $(k+1)$'th moment of this Rayleigh variable is given by
\begin{align}
    \mathbb{E}\left[|x_\text{in}|^{2(k+1)}\right] = P^{k+1} \Gamma\left[1+(k+1)\right] = P^{k+1} (k+1)!,
\end{align}
which can be replaced in \eqref{EyxProof} to find $g$ as
\begin{align}
    g = \frac{\mathbb{E}\left[y_\text{out}\bar{x}_\text{in}\right]}{\mathbb{E}\left[|x_\text{in}|^2\right]}=\sum_{k=0}^{L-1}a_{2k+1}(k+1)!~P^k.
\end{align}
In order to find $C_{\eta}$, we need to calculate $C_{y_\text{out}}$, which is given by
\begin{align} \label{CyyProof}
    \mathbb{E}\left[y_\text{out}\bar{y}_\text{out}\right] = \mathbb{E}\left[\sum_{m=0}^{L-1} a_{2m+1}x_\text{in}|x_\text{in}|^{2m}\sum_{n=0}^{L-1} \bar{a}_{2n+1}\bar{x}_\text{in}|x_\text{in}|^{2n}\right].
\end{align}
We can simplify it using a change of variable and considering the moments of a Rayleigh variable to get
\begin{align}
   \mathbb{E}\left[y_\text{out}\bar{y}_\text{out}\right] = \sum_{k=1}^{2L-1}\left(k!~P^k\sum_{i=1}^{k}a_{2i-1}\bar{a}_{2k-2i+1}\right).
\end{align}
\subsubsection{Proof of Lemma \ref{Reimann1}}
From \eqref{ReimannNorm2} and \eqref{IntegApprox}, we need to calculate the following integrals,
\begin{align}
    \frac{d}{2}\int_{r=0}^{R}\frac{1}{(d^2+r^2)^\frac{3}{2}}rdr, \\
    \frac{d^2}{8\pi}\int_{r=0}^{R}\frac{1}{(d^2+r^2)^3}rdr,
\end{align}
where we have simplified the phase part with a $2\pi$ factor before the integrals. By a change in variable as, $x=r^2$ and $rdr=\frac{1}{2}dx$, while using the following indefinite integral formulas,
\begin{align}
    \int\frac{1}{(d^2+x)^\frac{3}{2}}dx = \frac{-2}{\sqrt{d^2+x}}, \\
    \int\frac{1}{(d^2+x)^3}dx = \frac{-\frac{1}{2}}{(d^2+x)^2},
\end{align}
we can find the results in \eqref{Riemannform1} and \eqref{Riemannform2}, and a simple factorization gives $\eqref{Riemannform3}$.

\section{Algorithms}\label{AlgsAppendix}
In this Section, we provide algorithms to showcase that the optimization problems formulated in the paper can be solved numerically with limited computation complexity. Algorithm \ref{alg:alg2} provides a simple low-complexity numerical method to solve \eqref{sndrDensity}, where the accuracy of the solution depends on the number of iterations, $T$.
\begin{algorithm}[h]
	\caption{Numerical Method to solve \eqref{sndrDensity}}\label{alg:alg2}
	\begin{algorithmic}
		\STATE 
		\STATE\text{Define:}~
		$R_\mathrm{ub}=$ \eqref{Rub},~
		$R_\mathrm{lb}=$ \eqref{Rlb},~ 
		$\Delta=R_\mathrm{ub}-R_\mathrm{lb}$,~
		$R_\text{opt}=R_\mathrm{lb}$,~$R_\text{new}=R_\mathrm{lb}$,~
            $T \in \mathcal{N}$ where $T>>1$.
		\vspace*{3pt}
		\STATE {\text{\textbf{for}}}~$t=1$~\text{to}~$T$:
		\STATE \hspace{0.5cm} Calculate $\gamma$ from \eqref{sndrDensity} for $R=R_\text{new}$
		\STATE \hspace{0.5cm}\textbf{if} $\gamma \geq \gamma_0$:
		\STATE \hspace{1cm}$R_\text{opt}=R_\text{new}$
		\STATE \hspace{1cm}\textbf{break}
		\STATE \hspace{0.5cm}\textbf{else}:
		\STATE \hspace{1cm}$R_\text{new}=R_\text{lb}+\frac{t}{T}\Delta$
		\STATE \hspace{0.5cm}\textbf{endif}
		\STATE {\text{\textbf{endfor}}}
	\end{algorithmic}
\end{algorithm}

Algorithm \ref{alg:alg1} can be used to find a sub-optimal solution to the optimization problem \eqref{rRoptprob1}. By selecting a sufficiently large value for $T$, the sub-optimal solution from this algorithm can approach the optimal solution.
\begin{algorithm}[h]
	\caption{Sub-Optimal Solution to Problem \eqref{rRoptprob1}}\label{alg:alg1}
	\begin{algorithmic}
		\STATE 
		\STATE {\text{Define:~}}$\tilde{R}=NA/\pi$,~$r_0=0$,~$R_0=\sqrt{N_\text{max}A/\pi}$,~$\gamma_0=$~\eqref{SNDRonlyR} with $R=R_0$,~$r_{\text{opt}}=0$,~$R_{\text{opt}}=R_0$, $\gamma_\text{opt}=\gamma_0$,~$T \in \mathcal{N}$ where $T>>1$.
		\vspace*{3pt}
		\STATE {\text{\textbf{for}}}~$t=1$~\text{to}~$T$:
		\vspace*{3pt}
		\STATE \hspace{0.5cm}$r_\text{new}=\frac{t}{T}\sqrt{\tilde{R}^2-N_\text{max}A/pi}$,
		\vspace*{3pt}
		\STATE \hspace{0.5cm}$R_\text{new}=\sqrt{N_\text{max}A/\pi+r_\text{opt}^2}$,
		\vspace*{3pt}
		\STATE \hspace{0.5cm}$\gamma_\text{new}=$~\eqref{rRoptprob1} with $r=r_\text{new}$ and $R=R_\text{new}$,
		\vspace*{3pt}
		\STATE \hspace{0.5cm}\textbf{if} $\gamma < \gamma_\text{opt}$:
		\STATE \hspace{1cm}\textbf{break}
		\STATE \hspace{0.5cm}\textbf{else}:
		\STATE \hspace{1cm}$r_\text{opt}=r_\text{new}$,
		\STATE \hspace{1cm}$R_\text{opt}=R_\text{new}$,
		\STATE \hspace{1cm}$\gamma_\text{opt}=\gamma_\text{new}$,
		\STATE \hspace{0.5cm}\textbf{endif}
		\vspace*{3pt}
		\STATE {\text{\textbf{endfor}}}
	\end{algorithmic}
\end{algorithm}
 }

\bibliographystyle{IEEEtran}
\bibliography{IEEEabrv,Refs}

\end{document}